\documentclass[electronic]{vgtc}             

\ifpdf
  \pdfoutput=1\relax                   
  \usepackage{graphicx}                
  \DeclareGraphicsExtensions{.pdf,.png,.jpg,.jpeg} 
\else
  \usepackage{graphicx}                
  \DeclareGraphicsExtensions{.eps}     
\fi%

\graphicspath{{figures/}{pictures/}{images/}{./}} 

\usepackage{microtype}                 
\PassOptionsToPackage{warn}{textcomp}  
\usepackage{textcomp}                  
\usepackage{mathptmx}                  
\usepackage{times}                     
\usepackage{cite}                      
\usepackage{tabu}                      
\usepackage{booktabs}                  

\onlineid{29}

\vgtccategory{Research}

\vgtcinsertpkg

\title{Differentiable Subdivision Surface Fitting}

\author{Tianhao Xie\thanks{e-mail: tianhao.xie@mail.concordia.ca}\\ %
        \scriptsize Concordia University %
     }

\teaser{
  \includegraphics[width=\textwidth]{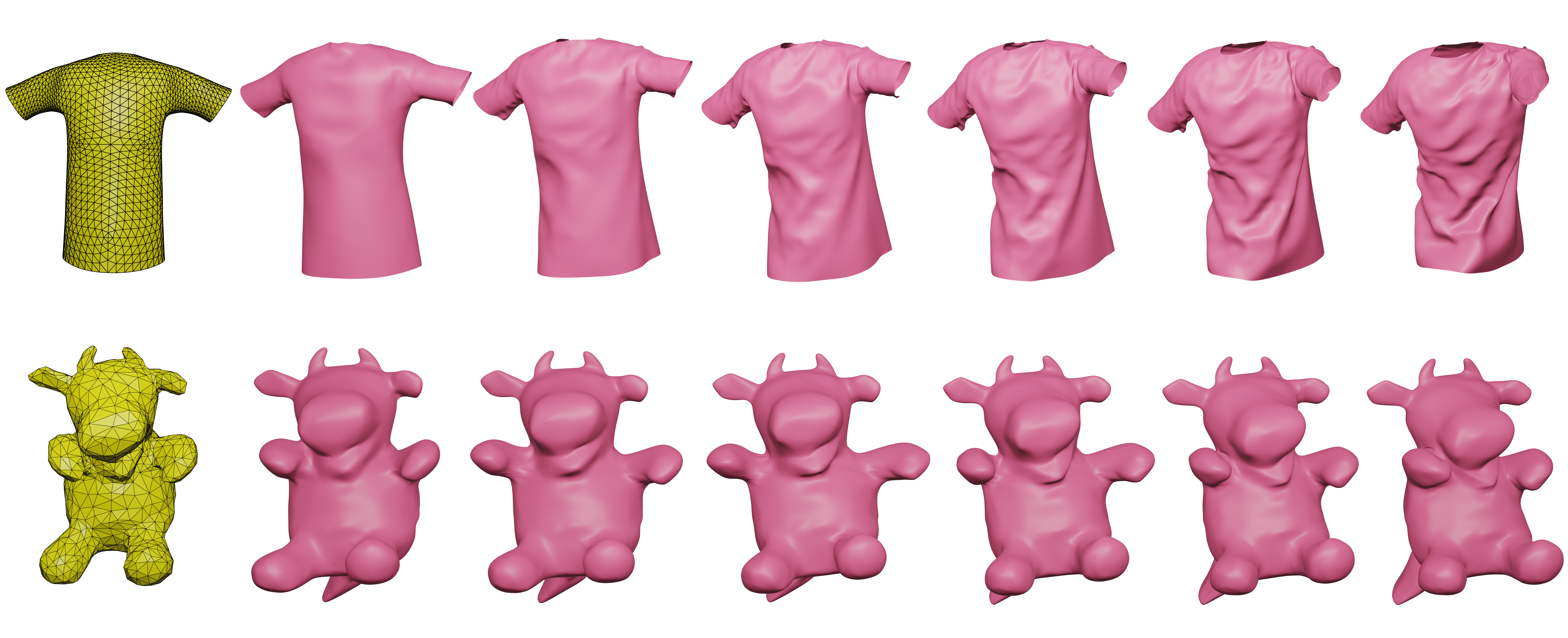}
  \caption{Fitting subdivision surface to spatial-temporal sequences}
}

\abstract{In this paper we present a powerful differentiable surface fitting technique to derive a compact surface representation for a given dense point cloud or mesh, with application in the domains of graphics and CAD/CAM. We have chosen the Loop subdivision surface, which in the limit yields the smooth surface underlying the point cloud, and can handle complex surface topology better than other popular compact representations, such as NURBS. The principal idea is to fit the Loop subdivision surface not directly to the point cloud, but to the IMLS (implicit moving least squares) surface defined over the point cloud. As both Loop subdivision and IMLS have analytical expressions, we are able to formulate the problem as an unconstrained minimization problem of a completely differentiable function that can be solved with standard numerical solvers. Differentiability enables us to integrate the subdivision surface into any deep learning method for point clouds or meshes. We demonstrate the versatility and potential of this approach by using it in conjunction with a differentiable renderer to robustly reconstruct compact surface representations of spatial-temporal sequences of dense meshes.}

\begin{document}


\maketitle
\section{Introduction} \label{Introduction}

Three-dimensional geometric data has many representations depending on the context in which it is used.
One such classification is by the number of parameters or control variables needed to define a 3D surface.
One extreme case would be to use raw geometric data obtained from 3D acquisition in the wild using 3D sensors or photogrammetric techniques; this is usually represented as a large and unstructured point cloud, where the sampled points (parameters) are noisy and the geometry is often incomplete. 
At the other end of the spectrum, for simulation, CAD, shape optimization, animation, and other modeling and analysis applications, a compact and precise representation is desired; examples are implicit algebraic surfaces~\cite{Bajaj92algebraicsurfaces}, bi-parametric surfaces using splines and NURBS~\cite{SplinesBookBarsky, PiegTill96} or subdivision surfaces~\cite{derose1998subdivision, montes2020computational}.
These compact surface representations use a relatively small number of control variables which are adjusted to yield a desired surface. 
In between these two extremes are a wide range of representations, from point-based surface definitions such as Moving Least Squares (MLS) surfaces, signed distance functions to standard polygonal meshes endowed with additional properties such as texture, normal and displacement maps.
Conversion between these various representations is fundamental to any geometric processing pipeline, and many have been developed over the years~\cite{berger2017survey}. 

Among all these representations, subdivision surfaces are particularly appealing to many high level applications such as surface optimization and analysis, simulation, modeling, and animation~\cite{derose1998subdivision}: not only they are very compact, they do not require explicit NURBS patch decomposition and alignment as NURBS do~\cite{sharma2020parsenet}, which makes them ideal to use for fitting more complex surface topology. A subdivision surface is represented by a compact polygonal mesh which gets subdivided by introducing new vertices, using, for example, Loop subdivision formulation~\cite{stam1998evaluation}; in the limit, this subdivision process leads to a smooth shape. With the ubiquity of colour and depth sensing, dense point clouds with very large number of points have become very easy to acquire, but are difficult to use for a number of reasons, including size, ambiguity of the underlying surface, shape editing difficulties, etc. Fitting a compact representation, such as a subdivision surface, would address a number of these problems. 

\begin{figure}[t]
  \centering
  \includegraphics[width=0.45\textwidth]{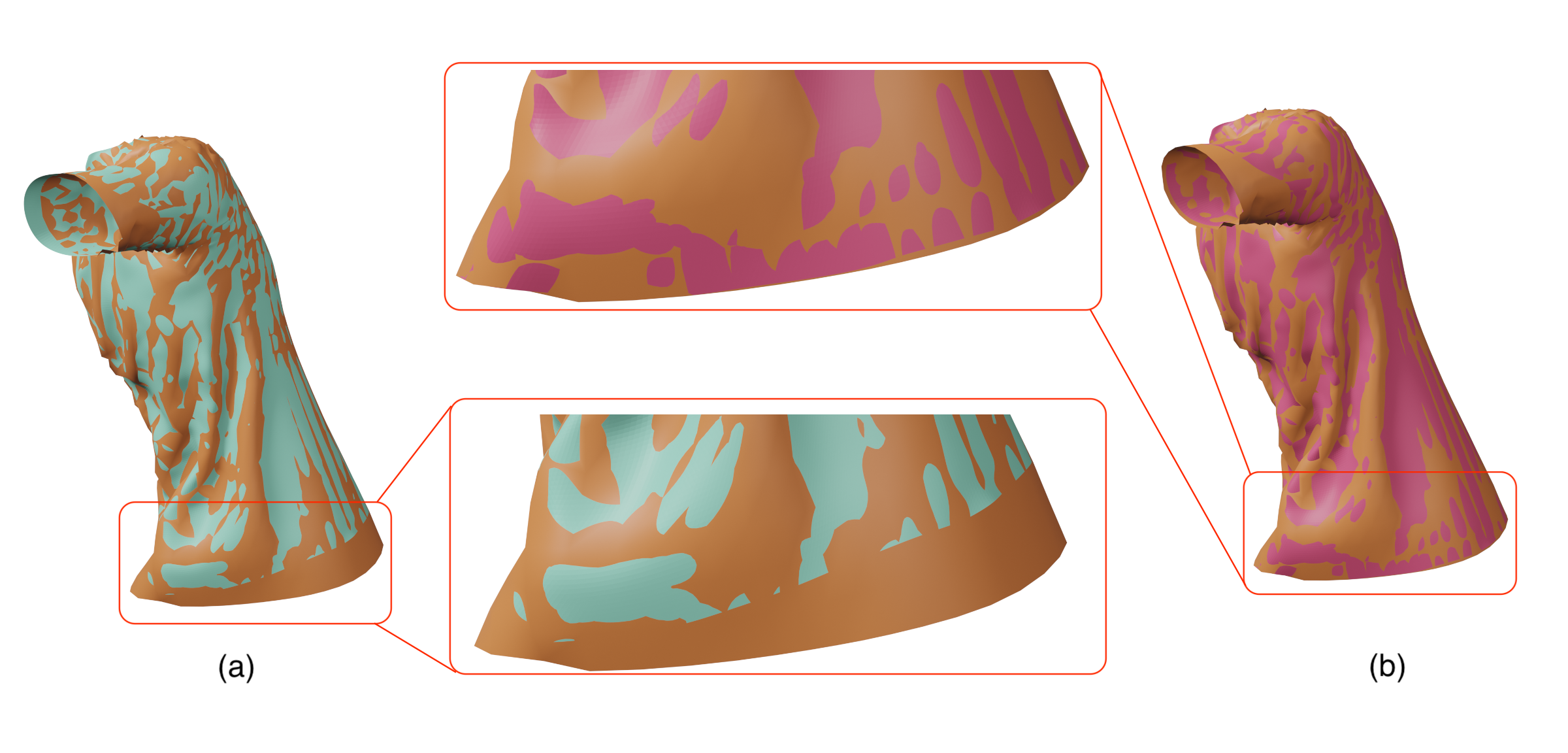}
  \caption{Comparison between the result of T-shirt data fitting. Brown color is the target. Green (a) using the geometric IMLS fit. Red (b) combining the geometric IMLS fit together with the image loss from the differential renderer. Note the drift in (a) at the bottom of the T-Shirt.}
  \label{fig:compare}
\end{figure}

Fitting subdivision surfaces to point clouds has major challenges. For a start, they require an initial control mesh that is capable of representing the desired surface topology in the limit. This is difficult to derive and consequently it is usually guessed, unless a mesh is extracted by other means. Existing fitting methods~\cite{estellers2018robust,cheng2004fitting,marinov2005optimization} rely on an optimization function that uses iterative point-to-point correspondences. This optimization function is non-differentiable, is not robust to noise and outliers, and also tends to fail if the initial guess of the control mesh is too far from the solution, especially in the tangential direction (see Figure~\ref{fig:compare}).  
The reason is that this point-to-point fitting strategy is not only non-differentiable but also rigid and does not easily allow for optimization along the tangent space. Optimization along the tangent space is particularly important especially when fitting to spatial-temporal data where we must fit a fixed topology template to spatially and temporally changing point clouds, which are extensively used in computer animation~\cite{mendhurwar2020system,popa2010globally}.

Some of these challenges conceptually stem simply from the fact that the gap between these representations is too large: whereas the subdivision surface represents a smooth continuous surface in a very compact way, the point cloud is only a collection of points and associated surface normals, with no other ordering or structure. 

Our principal idea in this work is to help the fitting by bridging this gap using an intermediate representation such as the implicit moving least square (IMLS)~\cite{IMLSKolluri08} surface. Thus, instead of fitting the subdivision surface directly to the point cloud, we fit it to the IMLS surface defined over the point cloud.  This bridging approach has several significant advantages. The IMLS surface plays the important role of an initial fairing operator over the point cloud. It defines an elegant and robust analytical distance function that replaces the traditional point to point distance used in previous methods, and one that naturally allows for sliding in the tangent space, making it ideal for both static and spatial-temporal surface fitting.

Our major contribution in this work is a complete pipeline for fitting a Loop subdivision surface to a dense point cloud or mesh using the IMLS surface as an intermediate representation. As both the Loop subdivision surface as well as the IMLS have analytical expressions, we are able to formulate the problem as an unconstrained minimization problem of a completely differentiable function that can be solved with standard numerical solvers. Furthermore, this differentiable surface fitting provides us with unique capability to integrate the compact subdivision surface representation of a point cloud into any deep learning method. We demonstrate the versatility and potential of this approach by using it in conjunction with a differentiable renderer to robustly reconstruct compact representations of spatial-temporal surface sequences.

\section{Related work} \label{Related}

The subdivision process defines a smooth curve or surface as the limit of a sequence of mesh refinement steps starting from a control mesh. This makes the final surface to be controlled by the small number of control vertices in the starting mesh, thus  resulting in a very compact surface representation.
Several subdivision schemes have been developed over the years and widely used in different applications\cite{catmull1978recursively, doo1978behaviour, derose1998subdivision, Loop:1987,Liu2021MLS}.
In particular, Loop subdivision is a subdivision scheme based on quartic box spline on triangular meshes~\cite{Loop:1987}. It is guaranteed that, in the limit, the subdivision surface has ${C^2}$ continuity in regular vertices (degree 6) and ${C^1}$ continuity in irregular vertices. In 1998, Jos Stam developed an analytical evaluation method of Loop subdivision \cite{stam1998evaluation}, which was based on conversion from Box splines to B-Nets \cite{lai1992fortran}. 
This analytical and differentiable evaluation makes this scheme ideal for differentiable shape optimization and we will use it in our novel subdivision fitting pipeline.





\subsection{Fitting subdivision surface to target shape}
It is a common task to fit a smooth surface representation to a target shape in computer graphics. One typical solution for this task is to fit a piecewise smooth surface to the target, such as a B-spline surface or a subdivision surface. Considerable work has been done on fitting B-splines to point clouds by squared distance minimization\cite{wang2006fitting,zheng2012fast}. Since our focus in this work is on fitting subdivision surfaces, we will limit our discussion of related work primarily to subdivision surface fitting.  

 Hoppe \emph{et al.}\cite{hoppe1994piecewise} and Lavoue \emph{et al.}\cite{lavoue2005subdivision} fit subdivision surfaces to CAD models by minimizing the squared distance energy. 
 Litke \emph{et al.}\cite{litke2001fitting} used quasi-interpolation to fit Catmull-Clark subdivision surface to a given shape within a prescribed tolerance. Ma \emph{et al.}\cite{Ma} described a method to fit a Loop subdivision surface to a dense triangular mesh by linear least square fitting.

The geometric data captured in the wild is almost always in the form of an unstructured point cloud, with noise, outliers, and missing geometry.
A large body of work has focused on fitting subdivision surfaces to point clouds. data\cite{estellers2018robust,cheng2004fitting,marinov2005optimization,mendhurwar2020system}. Cheng \emph{et al.}\cite{cheng2004fitting} fit the subdivision surface by iteratively minimizing a quadratic approximant of the squared distance function of a target shape. Their approach first samples  points on the Loop subdivision surface based on a method by Stam\cite{stam1998evaluation}. Then, they solve a linear system of the control mesh variables to minimize the squared distance between the sample points and target shape. 
Marinov \emph{et al.}\cite{marinov2005optimization} introduced an algorithm based on exact closest point search on Loop surfaces which combines Newton iteration and non-linear minimization. 
In more recent research, Esteller \emph{et al.}\cite{estellers2018robust} used second-order approximation of the squared distance function and the tangent space alignment to achieve robust fitting of subdivision surface for shape analysis. Similar to methods in \cite{cheng2004fitting} and \cite{marinov2005optimization}, Esteller \emph{et al.} also sampled the points on the subdivision surface to establish the error function -- error between the subdivision surface and the target shape. These methods need to solve a sequence of constrained least-squares problems to minimize the error function. The method in \cite{ilic2006using} could be optimized by gradient-descent method. However, instead of fitting the limit surface, they could only fit a specific level of subdivision surface to the target shape. 
In contrast to many of these methods, our proposed solution frames the fitting problem as an optimization of a completely differentiable function that can be solved using standard differentiable optimization methods.

Some learning-based methods to fit a surface to a target shape have also been previously proposed. 
Most of these approaches fit parametric polynomial surfaces of some form to point clouds. Yumer and Kara used a neural network to generate NURBS from input point sets\cite{yumer2012surface}. DeepFit incorporated a neural network to learn point-wise weights for weighted least squares polynomial surface fitting\cite{ben2020deepfit}. Sharma \emph{et al.} described a method using neural networks to fit B-spline patches to input point cloud data\cite{sharma2020parsenet}. Our fitting method is not deep learning based; however, being differentiable, it can be used to bridge the gap between deep learning-based methods in a 3D domain and traditional subdivision surface techniques. 

\subsection{Spatio-temporal surface reconstruction}
Reconstructing representations for time-varying 3D data is a common problem in graphics animation and simulation. A common approach is to fit a template mesh to the consecutive time-series point cloud or mesh. This is used to reconstruct coherent dynamic geometry from time-varying point clouds captured by real-time 3D scanning techniques. One widely used method is to reconstruct meshes for all frames first and then to fit a template mesh to all reconstructed meshes\cite{kahler2002head,allen2002articulated,sumner2004deformation,stoll2006template}. These methods always need additional markers or
landmarks which must be specified by the users. Another method is to generate a template from first frame and then fit the template directly to the remaining frames\cite{shinya2004unifying,sussmuth2008reconstructing}. In \cite{sussmuth2008reconstructing}, Sussmuth \emph{et al.} followed the Multi-level Partition of Unity (MPU) Implicits approach to reconstruct the implicit function that approximates the time-varying surface defined by the time-varying point cloud and used the As-Rigid-As-Possible constraint to the moving of the points. When comparing this approach to our method, 1) it does not fit a subdivision surface to the 4D data and thus the final resulting surface was not smooth; 2) unlike our distance field energy, they used an implicit function to represent the point cloud surface, which must be optimized by solving a sequence of least squares problems.

A few other methods perform template-free reconstruction~\cite{popa2010globally,mfoggp_dyn_reg_07,sharf2008space}. 
In \cite{mfoggp_dyn_reg_07}, Mitra \emph{et al.} directly compute the motion of the scanned object in all frames and estimate the time-deforming object by kinematic properties. In \cite{sharf2008space}, Sharf \emph{et al.} used a space-time solid incompressible flow prior to reconstruct moving and deforming objects from point data. 
In ~\cite{popa2010globally} a template is constructed gradually by mapping consecutive frames in a pyramidal fashion.
In \cite{wand2009efficient}, Wand \emph{et al.} reconstructed 3D scanner data by pairwise scanning alignment. Tevs \emph{et al.} \cite{tevs2012animation} introduced Animation Cartography, an intrinsic reconstruction of shape and motion, based on robust estimation of dense correspondences under topological noise and landmark tracking in temporally coherent and incoherent data. In addition, there are also some real-time reconstruction methods for general objects, such as \cite{newcombe2015dynamicfusion,li2009robust,zollhofer2014real}. 
Our method described next is distinct from all the above, specifically in our formulation using the IMLS surface as an intermediate for fitting.

\begin{figure}[t]
  \centering
  \includegraphics[width=0.45\textwidth]{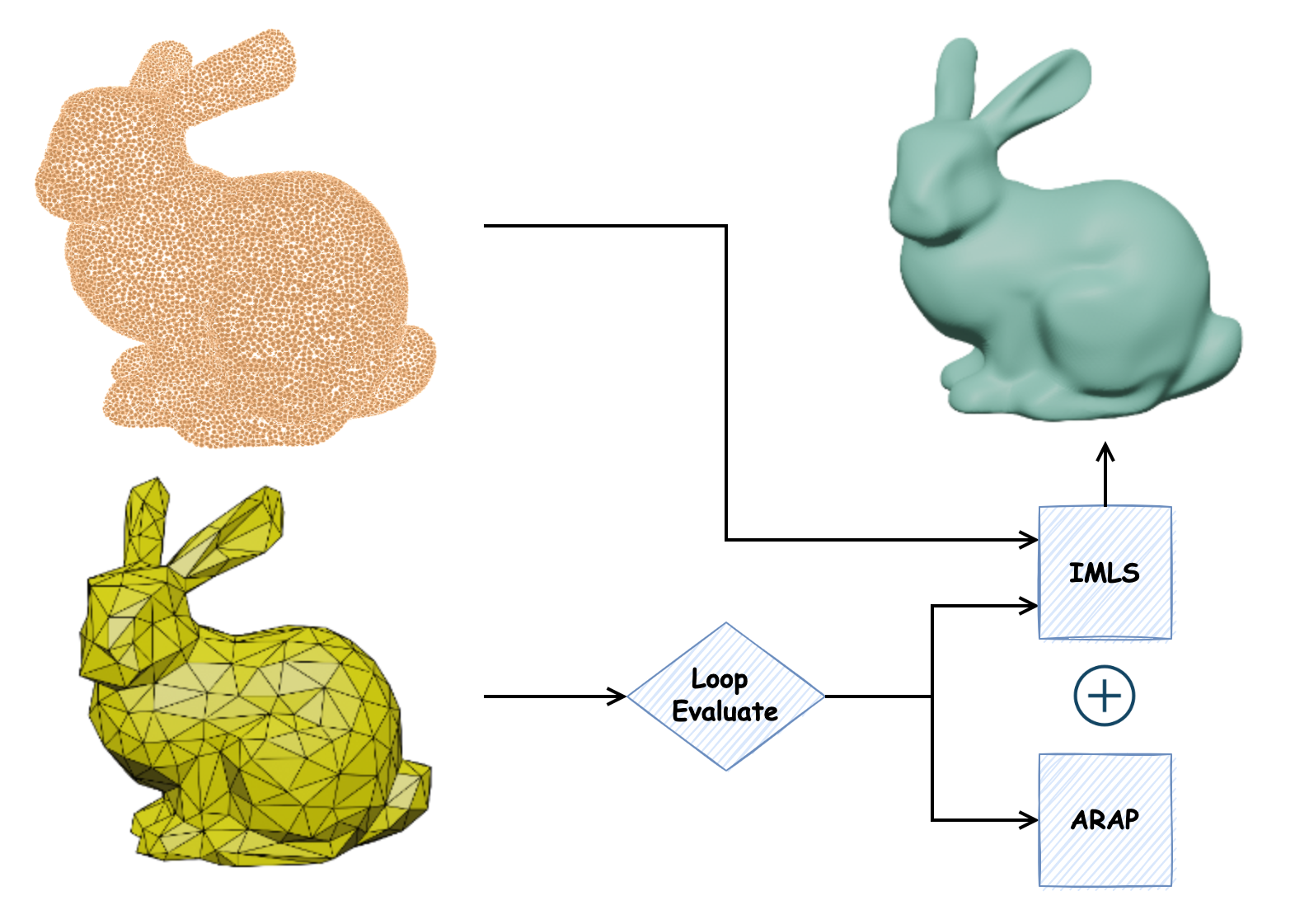}
  \caption{Overview of fitting subdivision surface to a static point cloud.}
  \label{fig:imls}
\end{figure}

\section{Method Overview} \label{Method}
The input to our pipeline is either a static target shape in the form of a point cloud $P$ or a temporal sequence of target shapes $S^i$ in the form of a set of triangular meshes. 
We note that we do not require the triangular meshes to have the same connectivity.
For the spatial-temporal case. we employ meshes as target shapes instead of point clouds only because there are currently no available reliable differentiable renderers for point clouds. And we need differentiable rendering since we want to combine it with our differentiable fitting to reconstruct spatial-temporal surfaces. But we would like to emphasize here that our method poses no conceptual limitations for using point clouds even for the spatial-temporal case.

\begin{figure}[t]
  \centering
  \includegraphics[width=0.45\textwidth]{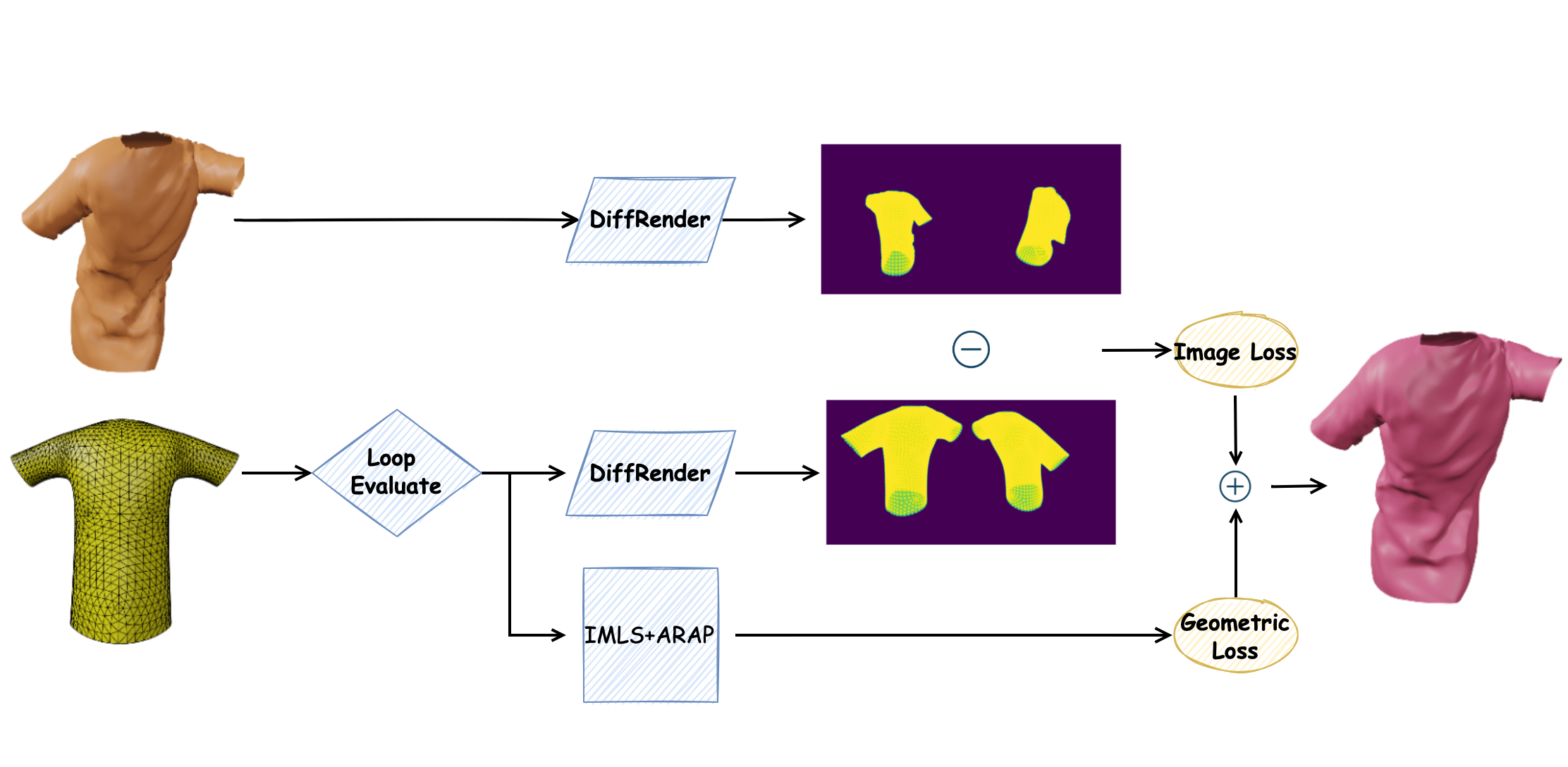}
  \caption{Overview of fitting subdivision surfaces to a spatial-temporal sequence by combining Implicit Moving Least Squares (IMLS) with differential rendering (DR) optimization}
  \label{fig:diff}
\end{figure}

The output of our method is a subdivision surface defined by a control mesh $M^0$. 
For the spatial-temporal fitting, the vertices of $M^0$ will have different 3D positions in each frame.
The overview of our method is presented in Figures~\ref{fig:imls} and \ref{fig:diff}.
From the point cloud we first create an initial control mesh.
We  optimize the vertex positions of this control mesh by minimizing the distance between the subdivision surface defined by this control mesh and the underlying IMLS surface defined by the target point cloud.

Similar to previous work, we compute this distance by sampling points on the subdivision surface, but in our formulation the sampled points are expressed as a differentiable analytical function of the control mesh and the distance function used is also a differentiable analytical function. This results in an unconstrained optimization problem of a differentiable analytical function that can be solved efficiently using standard off the shelf numerical methods. 
For the spatial-temporal case, we fit the subdivision surface defined by the control mesh iteratively to the temporally changing sequence of shapes, using the solution from one frame as a initial guess for the subsequent frame. 
Although this approach is popular and widely used~\cite{mendhurwar2020system}, it often fails due to accumulated drift arising from the inherently local nature of the geometric distance. 
Consequently, additional information is used to correct it, usually either in the form of boundary constraints~\cite{estellers2018robust} or other visual queues such as optical flow~\cite{bozic2020neural,popa2010globally}. 
Recently, with the development of differentiable renderers, rendered image difference metrics can be used to optimize shape~\cite{kanazawa2018learning}.
Adding the image difference loss from the differential renderer complements our pipeline, adding a global structure to our local geometric fit thus eliminating the drift and yielding a more accurate fit.

\subsection{Loop subdivision}
The first step in our process is to create the control mesh for the Loop subdivision surface $M^0(V^0, E^0)$.
Although the position of the template mesh vertices will be determined by our optimization, the number of vertices as well as the topology of this mesh must be determined a priori. 
For this, we compute an initial triangular mesh that fits the point cloud using existing meshing methods; we used Screened Poisson~\cite{kazhdan2013screened} method in MeshLab~\cite{LocalChapterEvents:ItalChap:ItalianChapConf2008:129-136}. 
We then simplify this triangulation using quadratic edge collapse\cite{garland1997surface} until we obtain the desired number of vertices requested by the user. 

\begin{figure}[t]
  \centering
  \includegraphics[width=0.3\textwidth]{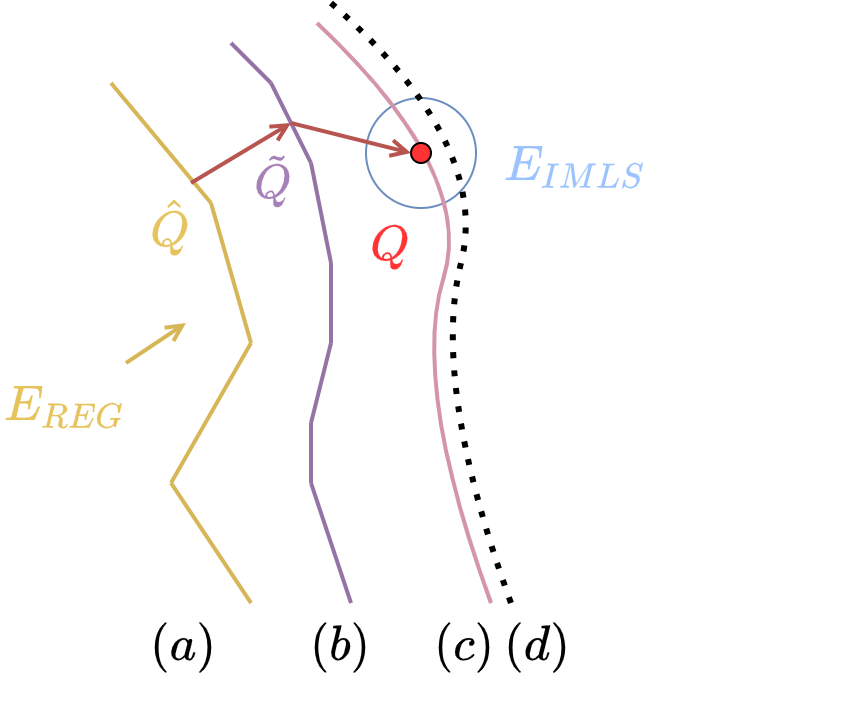}
  \caption{A schematic view of our optimization process. (a) Control mesh (${M^0}$). (b) Control mesh after one level of subdivision (${M^1}$). The vertices of this mesh are the Loop subdivision control points. (c) Loop subdivision surface. (d) Target point cloud. We optimize for ${M^0}$ by using an IMLS fitted to the point cloud and an ARAP regularizer on the control mesh ${M^0}$. }
  \label{fig:sketch}
\end{figure}

Although Loop subdivision can be evaluated iteratively, for optimization purposes it is desirable to have an analytical expression of the surface.
Jos Stam~\cite{stam1998evaluation} derived an analytical evaluation of the Loop subdivision surface of any point on the control mesh, but the scheme only works on the condition that no two adjacent vertices on the control mesh are extraordinary vertices (i.e. degree different from six). 
As it is very difficult to guarantee this condition especially when the control mesh has thousands of vertices, a solution is to apply just the first subdivision step obtaining a mesh $M^1(V^1, E^1)$. 
It can be easily proven that $M^1$ satisfies the above condition, however the number of vertices of $M^1$ are nearly four times as many as in $M^0$ making the representation far more verbose than desirable.
A key observation here is that even though the number of vertices of the mesh obtained after one level of subdivision $M^1$ is much larger, the added vertices can be computed analytically from the original mesh $M^0$ thus maintaining the same number of degrees of freedom in controlling the subdivision surface. 
Therefore, our control mesh for analytical evaluation of the subdivision surface is $M^1$, but we only optimize for the vertices of $M^0$.
This process is illustrated in Figure~\ref{fig:sketch}.
Given a point $\hat{Q}$ on the control mesh $M^0$, in order to compute its position on the final smooth 3D surface we first compute its position $\tilde{Q}$ on $M^1$ by using the Loop subdivision mask~\cite{Loop:1987}. Then after adjusting the triangle index and getting new barycentric coordinates we follow it by computing the position on the limit surface as per Stam~\cite{stam1998evaluation}.
This operator $L(\cdot)$ that maps the point $\hat{Q}$ on the control mesh to the point $Q$ onto the final subdivision surface is both analytical and differentiable.

\section{Subdivision Surface Fitting}
\subsection{Fitting a static model}
Given a point cloud ${P=\{P_i \}}$ with associated normals ${N=\{N_i \}}$, we fit our template control mesh $M^0(V^0, E^0)$ using the following optimization:

\begin{equation}
    \min_{V^0} \; E_{dist}(L(M^0,\hat{Q}))+\alpha \cdot E_{reg}(M^0, \bar{M^0}) 
\end{equation}
where ${E_{dist}(\cdot)}$ is the IMLS fit energy~\cite{oztireli2009feature} (eq.~\ref{imls_energy}), ${L(\cdot)}$ is the 3D position on the subdivision surface of a set of points $\hat{Q}$ sampled from the control mesh, $\bar{M_0}$ is the undeformed control mesh, ${E_{Reg}(\cdot)}$ is the ARAP regularizer~\cite{sorkine2007rigid} (eq.~\ref{arap_energy}) and ${\alpha}$ is the weight of the regularizer term. The overview of the fitting model is shown in Figure~\ref{fig:imls}.

\textbf{IMLS fit energy} Oztireli \emph{et al.} introduced an Implicit Moving Least Squares(IMLS) surface in \cite{oztireli2009feature}, which gave us a definition for the point cloud surface
\begin{equation}
\label{eq_imls1}
    f(x)={{\sum n_i^T(x-x_i) \phi_i(x)} \over {\sum \phi_i(x)}}
\end{equation}
where ${\phi}$ is a locally supported kernel function that vanishes beyond the cut-off distance ${h}$. ${h}$ is the radius we search for neighbor points and needs to be manually selected.
\begin{equation}
    \phi (r) = (1-{r^2 \over h^2})^4,
\end{equation}
We can use the implicit surface definition in equation ~\ref{eq_imls1} to derive a fit energy~\cite{montes2020computational}
\begin{equation} \label{imls_energy}
    E_{dist}={\sum_i ({{\sum_k N_k^T (Q_i-P_k)\phi(\| Q_i-P_k \|)} \over {\sum_k \phi(\|Q_i-P_k\|)}})^2} ,
\end{equation}
where ${P_k}$ and $N_k$ are the 3D positions and normals of points in the input point cloud (Figure~\ref{fig:sketch}d) and $Q_i$ are points on the subdivision surface sampled from the control mesh (Figure~\ref{fig:sketch}a-c).
For simplicity, in all our examples we only use the vertices of the control mesh, but we analyse the pros and cons of using more sampled points in the following sections and illustrated in figure~\ref{fig:lucy_compare}.

\begin{figure}[t]
  \centering
  \includegraphics[width=0.45\textwidth]{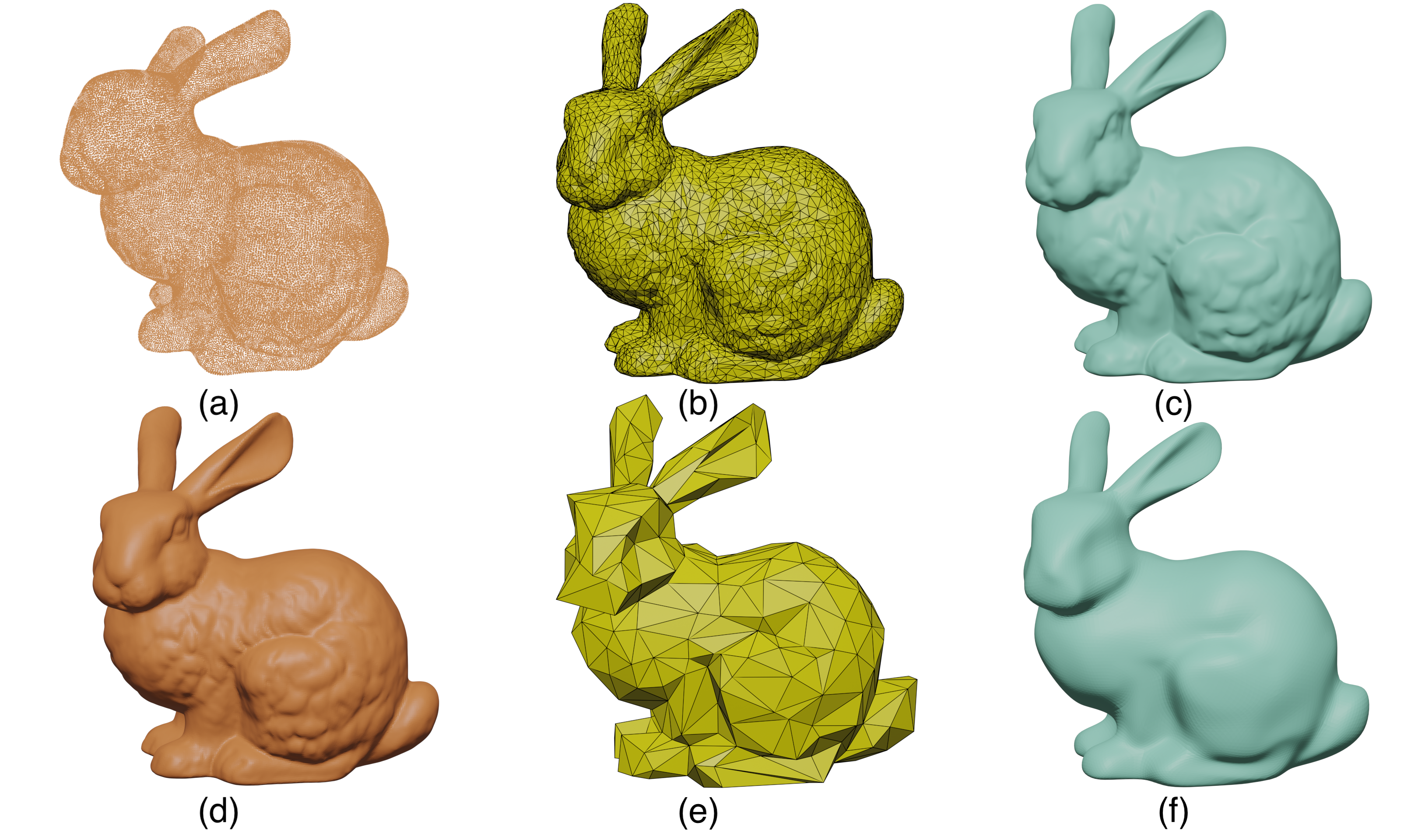}
  \caption{Stanford Bunny\cite{Stanford}:(a) Point cloud with 72,027 vertices. (b) Optimized control mesh with 4667 vertices. (c) Subdivision surface of (b). (d) Screened Poisson reconstructed mesh with 155,008 vertices. (e) Optimized control mesh with 314 vertices. (f) Subdivision surface of (e).}
  \label{fig:bunny}
\end{figure}

\begin{figure}[t]
  \centering
  \includegraphics[width=0.45\textwidth]{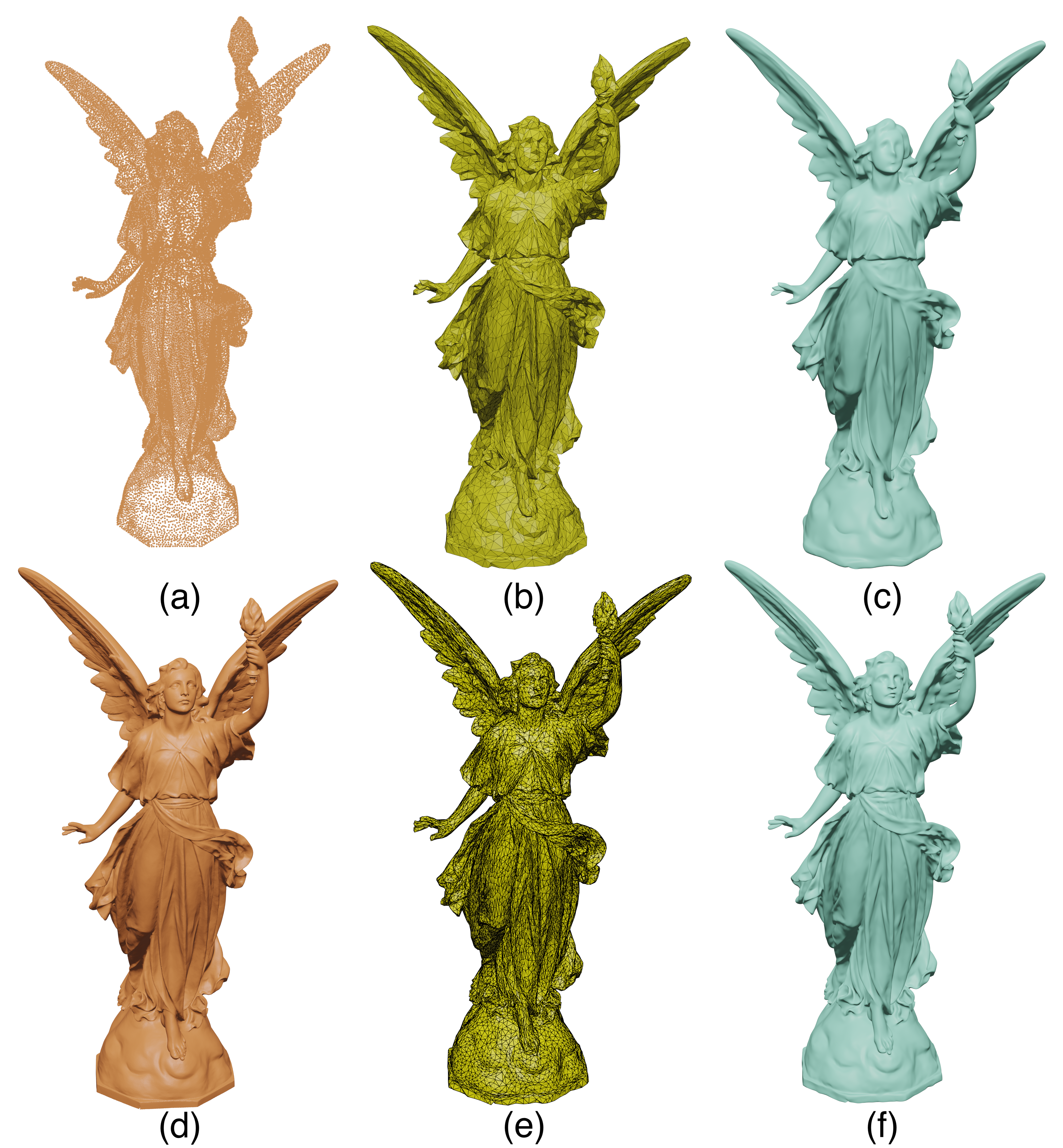}
  \caption{Stanford Lucy\cite{Stanford}:(a) Point cloud with 49,987 vertices. (b) Optimized control mesh with 8002 vertices. (c) Subdivision surface of (b). (d) Screened Poisson reconstructed mesh with 262,909 vertices. (e) Optimized control mesh with 20,002 vertices. (f) Subdivision surface of (e).}
  \label{fig:lucy}
\end{figure}



\textbf{Regularizer} We experimented with several reguralizers and the  As-Rigid-As-Possible (ARAP) regularizer~\cite{sorkine2007rigid} yields the best results. 
The ARAP regularizer does not penalize any isometric deformations allowing local rotations, but it penalizes local stretch.
More specifically:
\begin{equation} \label{arap_energy}
    E_{reg}=\sum_i \sum_{j \in N(i)} w_{ij}\|(V^0_i-V^0_j)-R_i(\bar{V^0_i}-\bar{V^0_j)}\|^2,
\end{equation}
where ${N(i)}$ is the set of vertices adjacent to ${V^0_i}$, $\bar{{V^0_i}}$ is the initial vertex position and ${R_i}$ is the local estimation rotation matrix for the one ring of vertices around vertex ${i}$. ${w_{ij}}$ is the standard cotangent Laplacian weight~\cite{meyer2003discrete}.
At every iteration ${R_i}$ can be computed analytically using SVD decomposition on the local co-variance matrix\cite{umeyama1991least}.


\textbf{Optimization} The optimization of the control mesh vertex positions ${V^0}$ is a non-linear optimization problem, which can be solved by using a non-linear solver, such as Google Ceres solver\cite{ceres-solver}. However, it will be slow if the size of the control mesh is large (i.e. thousands of vertices). In that case, we use gradient descent method, which is widely used in learning-based problem optimization. 

\subsection{Fitting a sptial-temporal model} \label{DR}
\textbf{Differentiable rendering}
The emergence of differentiable rendering (DR)~\cite{ravi2020accelerating,laine2020modular} paved the way for a new set of tools in 2D to 3D surface reconstruction.
 It allows 3D shape optimization and modeling from rendered 2D images~\cite{ravi2020accelerating,kato2020differentiable,kato2019learning,kanazawa2018learning}. 
 In image space, DR based optimization can give us a global loss energy when fitting to a mesh, which is complementary to our local geometric IMLS loss. Inspired by this, we introduce a new pipeline for fitting subdivision surface to spatial-temporal (4D) mesh data by combining our method with DR.

\textbf{Optimization}
Similar to the static case, given a control mesh $M^0$ and a sequence of spatial-temporal target meshes ${S^{i}}$, we are sequentially fitting the control mesh to each target mesh, using the solution of the current frame as an initial guess for the next one. 
Optimizing using only the geometric energy functionals described above leads to temporal drift as it can be seen in Figure~\ref{fig:compare} (a). 
Instead we add a image loss term that provides a global stabilization of the optimization, eliminating the drift as can be seen in Figure~\ref{fig:compare} (b). 

In every iteration's forward pass, we use the  DR  to render the target mesh in different camera positions $k$ which give us target images ${I_{TARGET}^{k}}$  
At same time, we use the same DR to render the limit surface of the template mesh which gives us predicted images $I_{PRED}^k$ in same camera positions as used for rendering the target images. 
Suppose the number of pixels for rendered images is ${N}$, we compute image loss ${l_{image}}$ by
\begin{equation}
    l_{image}={\sum_k {{(I_{PRED}^k - I_{TARGET}^k)^2} \over N}}.
\end{equation}
As for the geometric loss, we compute it by the same method for computing energy provided in section \ref{Method}. Thus, the total loss ${l_{total}}$ is
\begin{equation}
    l_{total}=E_{dist}(L(M^0),\hat{Q})+\alpha \cdot E_{reg}(M^0, \bar{M^0})+\beta \cdot l_{image}.
\end{equation}

In our implementation we use the DR available in PyTorch3D~\cite{ravi2020accelerating} and 
for the backward pass, we use the gradient descent method, Adam optimizer\cite{kingma2014adam}, to optimize the control mesh.


\section{Results and Discussion}
\subsection{Static surface fitting}
\begin{figure}[t]
  \centering
  \includegraphics[width=0.45\textwidth]{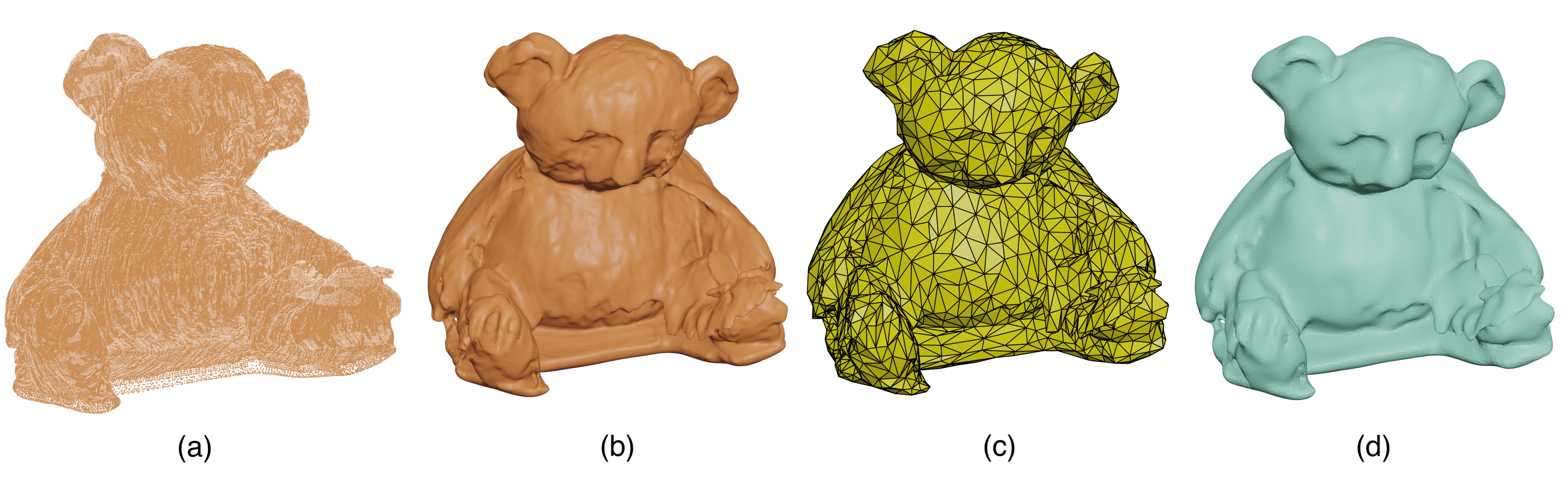}
  \caption{Kinect scanned koala:(a) Point cloud with 1,018,126 vertices. (b) Screened Poisson reconstructed mesh with 276,529 vertices. (c) Optimized control mesh with 2,465 vertices. (d) Subdivision surface of (c).}
  \label{fig:bear}
\end{figure}

We used our method to fit a number of synthetic models (the Stanford Bunny and Lucy, see Figures~\ref{fig:bunny} and ~\ref{fig:lucy} as well as a point cloud acquired using the Microsoft Azure Kinect device (a Koala toy) Figure~\ref{fig:bear}
The starting searching radius ${h_0}$ and weight of ARAP regularizer ${\alpha}$ were selected manually. We scaled the point clouds to a unit box before fitting, to increase the numerical stability of the optimization. For the Stanford Bunny and Lucy, we used ${h_0=0.0005}$. For the toy Koala, we used ${h_0=0.05}$. As for the ${\alpha}$, it depends on the noise level of the point cloud. When the point cloud is noisy, you need a bigger weight, such as 0.1. When the point cloud is very clean, ${\alpha}$ should be set to very small, such as 0.01. For the Stanford Bunny and Lucy, we set ${\alpha=0.01}$. For the toy Koala, we set ${\alpha=0.1}$. 

For the bunny(Figure~\ref{fig:bunny}) the original point cloud has $72,027$ vertices and the reconstructed mesh using Screened Poisson\cite{kazhdan2013screened} has $155,008$ vertices. We demonstrate two reconstructions.
The first one with a template mesh of $4667$ vertices (Figure~\ref{fig:bunny} (c)) that shows no visual difference to the original, but uses only around $3\%$ of the Screened Poisson reconstruction. 
The second one uses only $314$ vertices, or only $0.2\%$ of the Screened Poisson reconstruction (Figure~\ref{fig:bunny} (f)). While a number of details are lost, the main shape is still reconstructed fairly well. 

For the more detailed and complicated Lucy model (Figure~\ref{fig:lucy}), with only $3\%$ of the Screened Poisson reconstruction vertices, we could retain most of the intricate objects and folds. 

In Figure~\ref{fig:bear} we show the reconstruction of a koala toy. 
The physical scanned model is furry so while the original reconstruction is very detailed it also contained a lot of noise. With only $0.2\%$ of the original number of vertices and $0.8\%$ of Screened Poisson reconstruction vertices, we provide a reconstruction that retains the shape and many of the important details. 

\begin{figure}[t]
  \centering
  \includegraphics[width=0.45\textwidth]{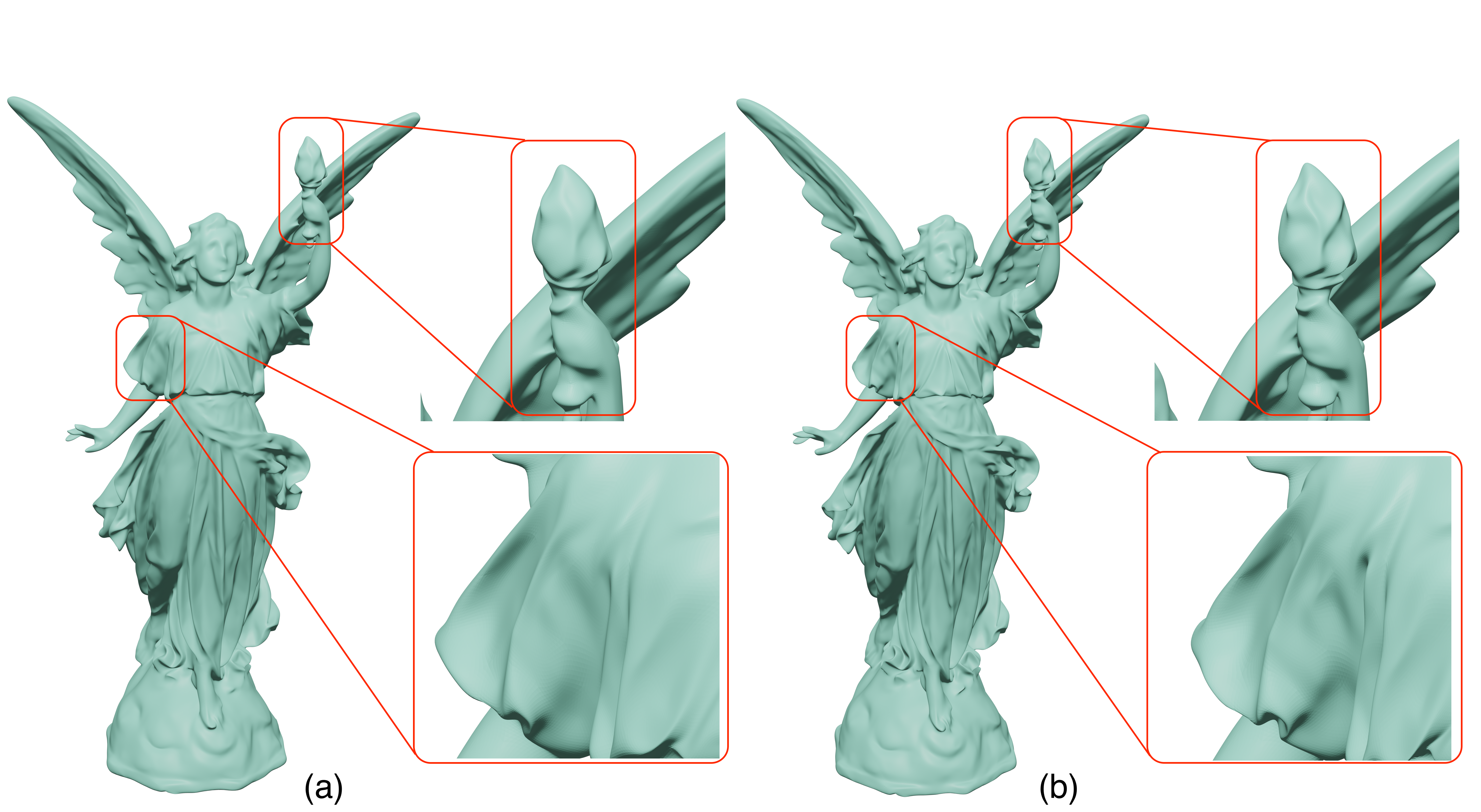}
  \caption{Comparison between (a) using only the control mesh vertices to compute the IMLS fit, and (b) using the vertices after one level of subdivision.}
  \label{fig:lucy_compare}
\end{figure}

\begin{figure*}[t]
  \centering
  \includegraphics[width=\textwidth]{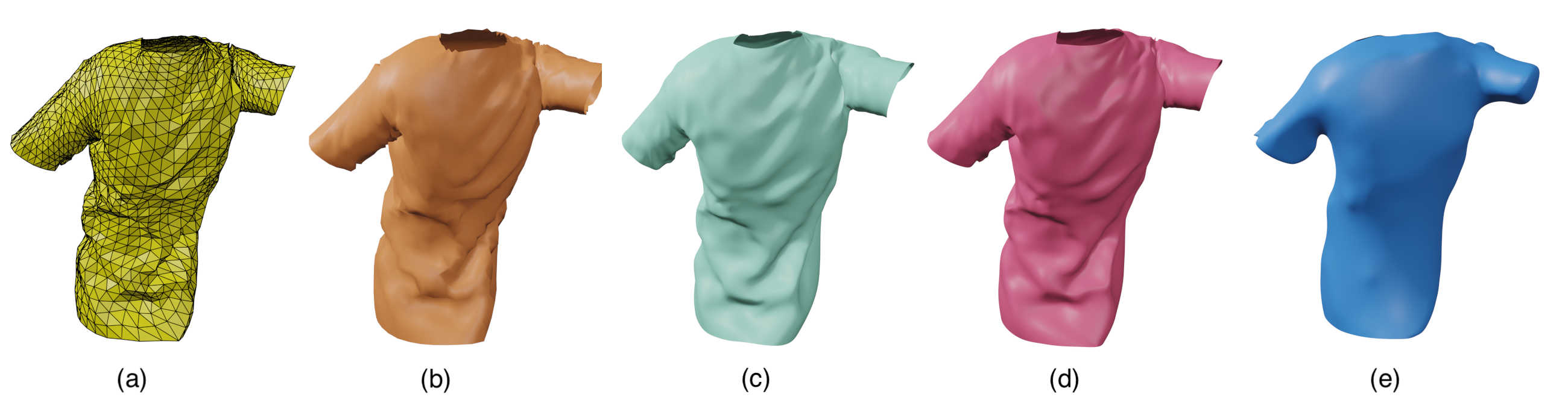}
  \caption{Fitting result for t-shirt simulation: (a) Optimized control mesh of using both IMLS and DR. (b) Simulation result from Blender\cite{Hess:2010:BFE:1893021}. (c) Fitting result by only IMLS energy(section \ref{Method}). (d) Fitting result by combining IMLS and DR(section \ref{DR}). (e) Fitting result by only DR.}
  \label{fig:t-shirt}
\end{figure*}

\begin{figure*}[!h]
  \centering
  \includegraphics[width=0.9\textwidth]{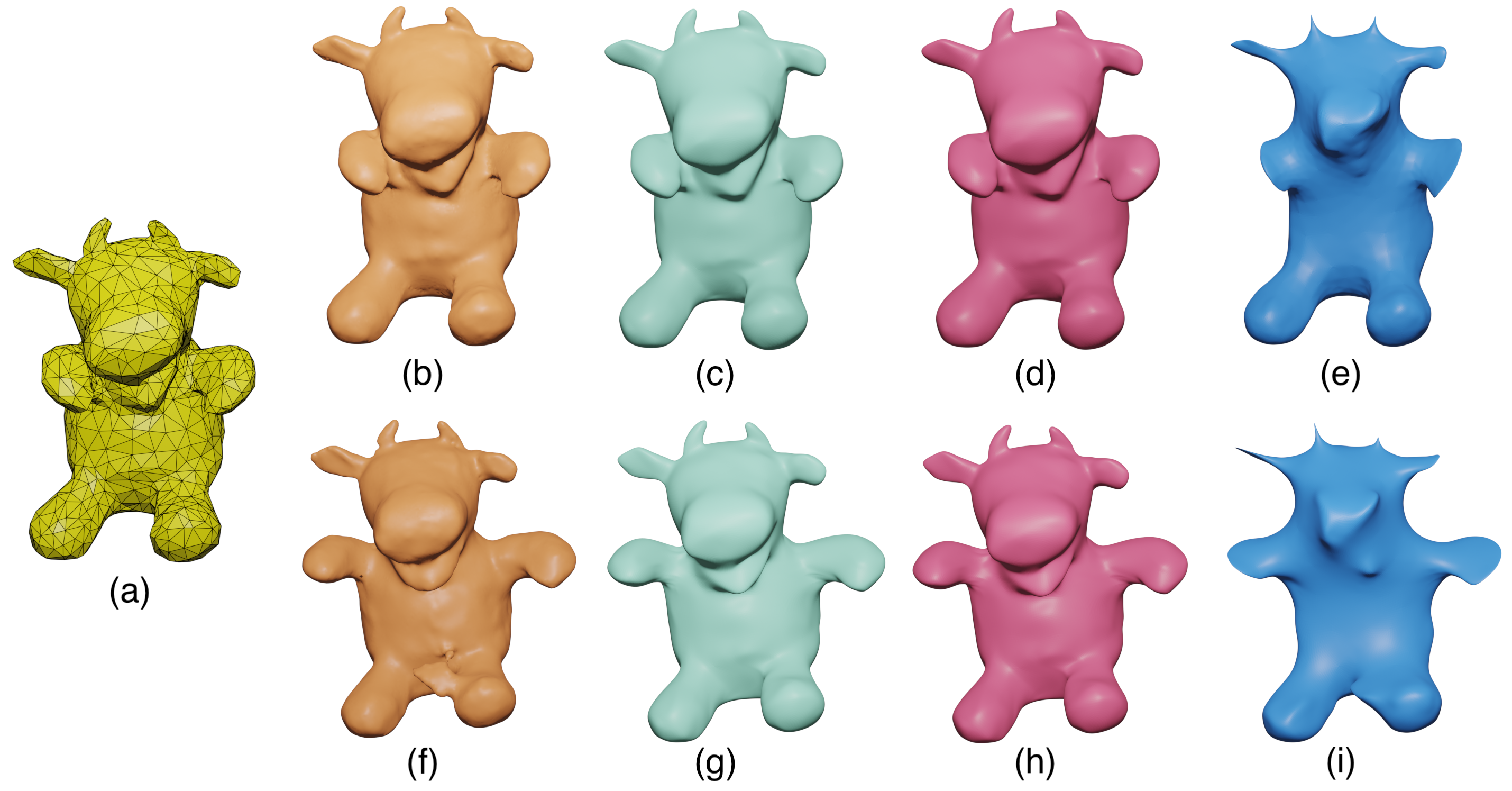}
  \caption{Fitting result for real scanned puppet. (a) Template control mesh with 1,252 vertices. (b) Reconstructed mesh with 123,234 vertices for start frame. (c)Fitted subdivision surface using IMLS energy(section \ref{Method} for start frame. (d) Fitted subdivision surface using combination of IMLS energy and DR(section \ref{DR}) for start frame. (e) Fitting result only using DR for start frame. (f) Reconstructed mesh with 123,234 vertices for end frame. (g) Fitted subdivision surface using IMLS energy for end frame. (h) Fitted subdivision surface using combination of IMLS energy and DR for end frame. (i) Fitting result only using DR for end frame.}
  \label{fig:puppet}
\end{figure*}

The performance of the IMLS distance depends on the number of sampled points on the subdivision surface that we use in the computation. 
By default in all our examples we only use the points in the control mesh. 
However, it is possible to select more samples. 
Figure~\ref{fig:lucy_compare} shows this trade off. 
Figure~\ref{fig:lucy_compare} (a) is the reconstruction of the Lucy model using only the vertices in the original control mesh. 
Figure~\ref{fig:lucy_compare} (b) is the reconstruction using the vertices obtained after one level of subdivision (i.e. four times more). 
The result is slightly improved, some areas contain more detail, but the optimization takes about three times as long. 


\subsection{Spatial-temporal fitting}
We tested our spatial-temporal method on two sequences: a synthetic sequence generated using a cloth simulation of a T-Shirt in Blender\cite{Hess:2010:BFE:1893021}, and a spatial-temporal capture of a cow toy using a multi-view stereo setup.
Both sequences have $30$ frames and 
in both cases we made a template from the first frame.
For the cloth sequence we used for simulation a mesh of $2000$ vertices that we randomly re-sampled in every frame to simulate a real capture to $100,000$ vertices (or 2\% of the total vertices). The template mesh has $2046$ vertices,
For the puppet sequence the target mesh has around $123,000$ vertices and the template mesh of $1252$ vertices (1\% of the total vertices).

The settings for the DR are adapted from the PyTorch3D\cite{ravi2020accelerating} tutorial. We used Soft Silhouette shader whose image size is ${256 \times 256}$, blur radius is ${log(1 / (1e^{-4} - 1) * 1e^{-4})}$ and faces per pixel is 100. When rendering the target shape, we had 20 different camera views in total. However, in every iteration, we only randomly select $2$ views to render the images of template to reduce unnecessary rendering time. 
The T-Shirt sequence has a lot of geometric details that is well preserved in the reconstruction. 
In contrast, the puppet sequence has less detail and in some cases some reconstruction artifacts (see Figure~\ref{fig:puppet} (f)) stay fixed in the reconstruction due to the continuity properties of the subdivision surfaces.

\begin{figure}[t]
  \centering
  \includegraphics[width=0.45\textwidth]{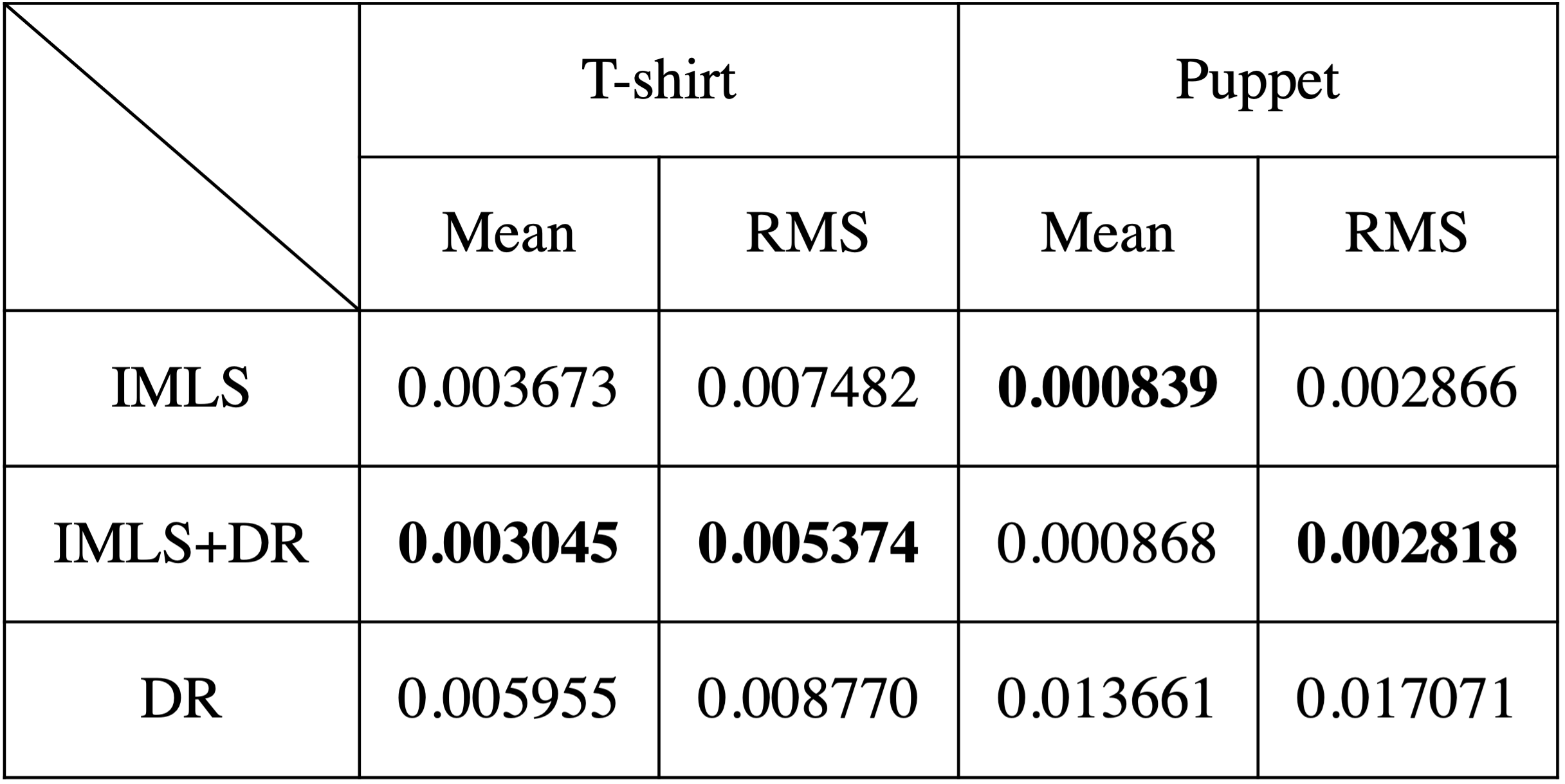}
  \caption{Hausdorff distance between fitting result and target shape(w.r.t bounding box diagonal)}
  \label{fig:table_hausdorff}
\end{figure}

\begin{figure}[t]
  \centering
  \includegraphics[width=0.45\textwidth]{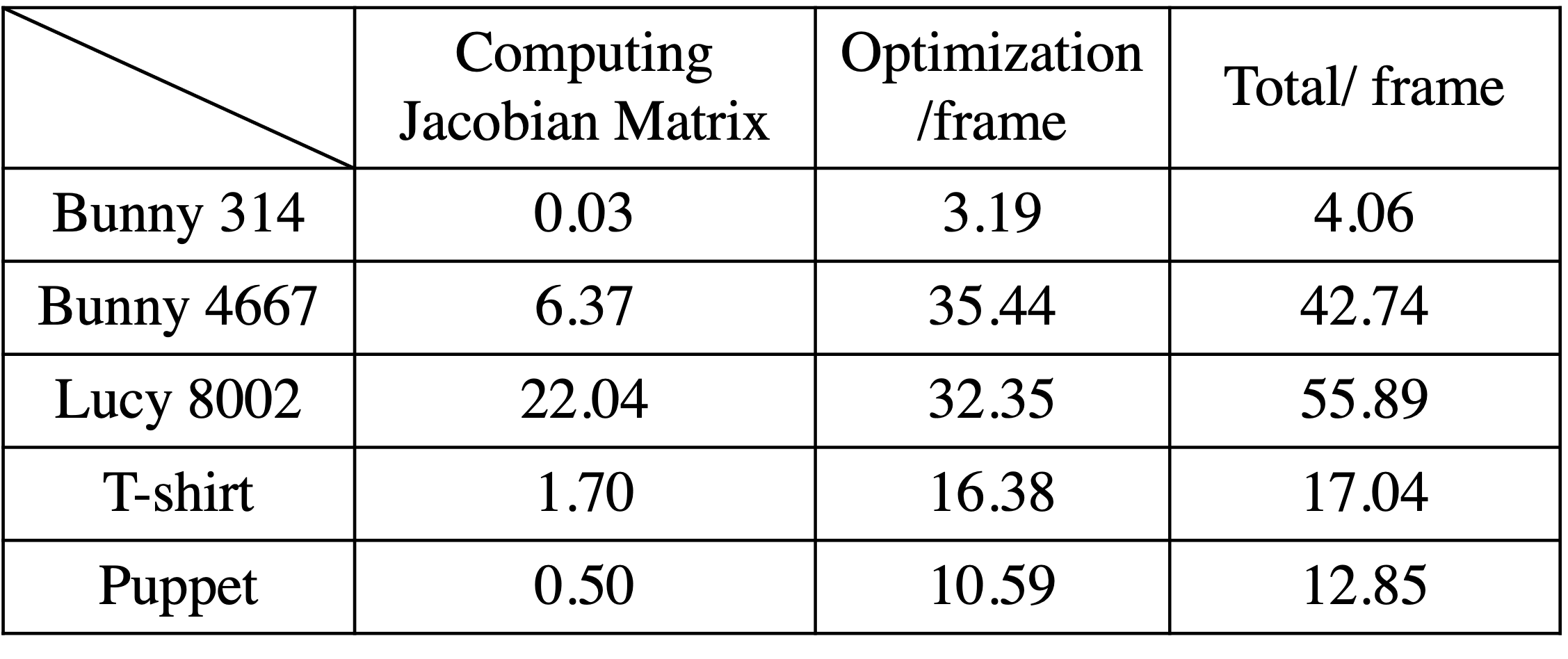}
  \caption{Running time (in seconds) for fitting subdivision surface to static model and fitting subdivision surface to spatial-temporal sequence(30 frames)}
  \label{fig:table_time}
\end{figure}

In Figures~\ref{fig:t-shirt} and ~\ref{fig:puppet} we compare the IMLS fitting scheme with the DR fitting scheme.
Using the DR fitting scheme by itself results in the loss of a lot of details: Figures~\ref{fig:t-shirt} (e), ~\ref{fig:puppet} (e), (i)
This is not unexpected as we only use the silhouette loss.
However, the geometric detail between IMLS and IMLS+DR is very similar (Figures ~\ref{fig:t-shirt} (c), (d), Figures ~\ref{fig:puppet} (c), (d), Figures ~\ref{fig:puppet} (g), (h)). The main gain from adding the DR term is the reduced drift (Figure~\ref{fig:compare}).
We also perform a quantitative evaluation using the Hausdorff distance between the target mesh and the subdivision surface. For the subdivision surface, we computed the Hausdorff distance using $3$ iterations of subdivision. Results are presented in Figure~\ref{fig:table_hausdorff}.
The combination of IMLS + DR largely outperforms either of them used separately.
Figure~\ref{fig:table_time} shows the execution time of our method. The code has been run on a computer with a CPU i9 12900 and GPU is rtx3060 ti.

\section{Conclusion, Limitations and Future work}
In this paper, we present a novel differentiable method to fit a Loop subdivision surface to a point cloud. 
Because our fitting method is differentiable, it can be easily integrated into deep learning-based methods for different applications.  
We demonstrate our method on several static point clouds as well as spatial-temporal shape sequences. 
The results show that our method does well in preserving surface detail while still being very compact, requiring only a small fraction (between 1\% and 3\%) of the data, in comparison to reconstruction methods such as Screened Poisson.

However, our method has some limitations.
The spatial-temporal reconstruction relies on a differential renderer and the ones currently available only support mesh format. Therefore, for the spatial-temporal examples, we had to reconstruct a triangular mesh from each static point cloud. 
Since the IMLS energy is based on nearest neighbor search, the optimization may fail when the distance between template control mesh and the target shape is too large. Thus, especially in the case of spatial-temporal examples, the frame-to-frame motion of the data must be relatively small.  

Since we focused on geometric fitting, we  selected an image loss based on silhouette only. In the future, it would be of interest to explore other image losses and point based differential renderers. 
In the future, we can also improve our method by using more accurate implicit surface reconstruction techniques from point-clouds such as the one proposed by Liu et al.~\cite{Liu2021MLS}.

\bibliographystyle{abbrv-doi}

\bibliography{template}

\begin{thebibliography}{10}

\bibitem{ceres-solver}
S.~Agarwal, K.~Mierle, and Others.
\newblock Ceres solver.
\newblock \url{http://ceres-solver.org}.

\bibitem{allen2002articulated}
B.~Allen, B.~Curless, and Z.~Popovi{\'c}.
\newblock Articulated body deformation from range scan data.
\newblock {\em ACM Transactions on Graphics (TOG)}, 21(3):612--619, 2002.

\bibitem{Bajaj92algebraicsurfaces}
C.~L. Bajaj.
\newblock Chapter 2: Surface fitting using implicit algebraic surface patches.
\newblock In H.~Hagen, ed., {\em Topics in Surface Modeling}, pp. 23--52.
  {SIAM}, 1992. doi: {{%
10\hspace{.1pt}\discretionary{.}{%
}{.}\hspace{.4pt}1137\discretionary{/}{%
}{/}1\hspace{.1pt}\discretionary{.}{%
}{.}\hspace{.4pt}9781611971644\hspace{.1pt}\discretionary{.}{%
}{.}\hspace{.4pt}ch2}}


\bibitem{SplinesBookBarsky}
R.~H. Bartels, J.~C. Beatty, and B.~A. Barsky.
\newblock {\em An Introduction to Splines for Use in Computer Graphics and
  Geometric Modeling}.
\newblock Morgan Kaufmann Publishers Inc., San Francisco, CA, USA, 1987.

\bibitem{ben2020deepfit}
Y.~Ben-Shabat and S.~Gould.
\newblock Deepfit: 3d surface fitting via neural network weighted least
  squares.
\newblock In {\em European Conference on Computer Vision}, pp. 20--34.
  Springer, 2020.

\bibitem{berger2017survey}
M.~Berger, A.~Tagliasacchi, L.~M. Seversky, P.~Alliez, G.~Guennebaud, J.~A.
  Levine, A.~Sharf, and C.~T. Silva.
\newblock A survey of surface reconstruction from point clouds.
\newblock In {\em Computer Graphics Forum}, vol.~36, pp. 301--329. Wiley Online
  Library, 2017.

\bibitem{bozic2020neural}
A.~Bozic, P.~Palafox, M.~Zollh{\"o}fer, A.~Dai, J.~Thies, and M.~Nie{\ss}ner.
\newblock Neural non-rigid tracking.
\newblock {\em Advances in Neural Information Processing Systems},
  33:18727--18737, 2020.

\bibitem{catmull1978recursively}
E.~Catmull and J.~Clark.
\newblock Recursively generated b-spline surfaces on arbitrary topological
  meshes.
\newblock {\em Computer-aided design}, 10(6):350--355, 1978.

\bibitem{cheng2004fitting}
K.-S. Cheng, W.~Wang, H.~Qin, K.-Y. Wong, H.~Yang, and Y.~Liu.
\newblock Fitting subdivision surfaces to unorganized point data using sdm.
\newblock In {\em 12th Pacific Conference on Computer Graphics and
  Applications, 2004. PG 2004. Proceedings.}, pp. 16--24. IEEE, 2004.

\bibitem{LocalChapterEvents:ItalChap:ItalianChapConf2008:129-136}
P.~Cignoni, M.~Callieri, M.~Corsini, M.~Dellepiane, F.~Ganovelli, and
  G.~Ranzuglia.
\newblock {MeshLab: an Open-Source Mesh Processing Tool}.
\newblock In V.~Scarano, R.~D. Chiara, and U.~Erra, eds., {\em Eurographics
  Italian Chapter Conference}. The Eurographics Association, 2008. doi: {{%
10\hspace{.1pt}\discretionary{.}{%
}{.}\hspace{.4pt}2312\discretionary{/}{%
}{/}LocalChapterEvents\discretionary{/}{%
}{/}ItalChap\discretionary{/}{%
}{/}ItalianChapConf2008\discretionary{/}{%
}{/}129\discretionary{%
}{-}{-}136}}


\bibitem{derose1998subdivision}
T.~DeRose, M.~Kass, and T.~Truong.
\newblock Subdivision surfaces in character animation.
\newblock In {\em Proceedings of the 25th annual conference on Computer
  graphics and interactive techniques}, pp. 85--94, 1998.

\bibitem{doo1978behaviour}
D.~Doo and M.~Sabin.
\newblock Behaviour of recursive division surfaces near extraordinary points.
\newblock {\em Computer-Aided Design}, 10(6):356--360, 1978.

\bibitem{estellers2018robust}
V.~Estellers, F.~Schmidt, and D.~Cremers.
\newblock Robust fitting of subdivision surfaces for smooth shape analysis.
\newblock In {\em 2018 International Conference on 3D Vision (3DV)}, pp.
  277--285. IEEE, 2018.

\bibitem{garland1997surface}
M.~Garland and P.~S. Heckbert.
\newblock Surface simplification using quadric error metrics.
\newblock In {\em Proceedings of the 24th annual conference on Computer
  graphics and interactive techniques}, pp. 209--216, 1997.

\bibitem{Hess:2010:BFE:1893021}
R.~Hess.
\newblock {\em Blender Foundations: The Essential Guide to Learning Blender
  2.6}.
\newblock Focal Press, 2010.

\bibitem{hoppe1994piecewise}
H.~Hoppe, T.~DeRose, T.~Duchamp, M.~Halstead, H.~Jin, J.~McDonald,
  J.~Schweitzer, and W.~Stuetzle.
\newblock Piecewise smooth surface reconstruction.
\newblock In {\em Proceedings of the 21st annual conference on Computer
  graphics and interactive techniques}, pp. 295--302, 1994.

\bibitem{ilic2006using}
S.~Ilic.
\newblock Using subdivision surfaces for 3-d reconstruction from noisy data.
\newblock In {\em Workshop on Image Registration in Deformable Environments
  (DEFORM)}, pp. 1--10. Citeseer, 2006.

\bibitem{kahler2002head}
K.~K{\"a}hler, J.~Haber, H.~Yamauchi, and H.-P. Seidel.
\newblock Head shop: Generating animated head models with anatomical structure.
\newblock In {\em Proceedings of the 2002 ACM SIGGRAPH/Eurographics symposium
  on Computer animation}, pp. 55--63, 2002.

\bibitem{kanazawa2018learning}
A.~Kanazawa, S.~Tulsiani, A.~A. Efros, and J.~Malik.
\newblock Learning category-specific mesh reconstruction from image
  collections.
\newblock In {\em Proceedings of the European Conference on Computer Vision
  (ECCV)}, pp. 371--386, 2018.

\bibitem{kato2020differentiable}
H.~Kato, D.~Beker, M.~Morariu, T.~Ando, T.~Matsuoka, W.~Kehl, and A.~Gaidon.
\newblock Differentiable rendering: A survey.
\newblock {\em arXiv preprint arXiv:2006.12057}, 2020.

\bibitem{kato2019learning}
H.~Kato and T.~Harada.
\newblock Learning view priors for single-view 3d reconstruction.
\newblock In {\em Proceedings of the IEEE/CVF Conference on Computer Vision and
  Pattern Recognition}, pp. 9778--9787, 2019.

\bibitem{kazhdan2013screened}
M.~Kazhdan and H.~Hoppe.
\newblock Screened poisson surface reconstruction.
\newblock {\em ACM Transactions on Graphics (ToG)}, 32(3):1--13, 2013.

\bibitem{kingma2014adam}
D.~P. Kingma and J.~Ba.
\newblock Adam: A method for stochastic optimization.
\newblock {\em arXiv preprint arXiv:1412.6980}, 2014.

\bibitem{IMLSKolluri08}
R.~K. Kolluri.
\newblock Provably good moving least squares.
\newblock {\em {ACM} Trans. Algorithms}, 4(2):18:1--18:25, 2008. doi: {{%
10\hspace{.1pt}\discretionary{.}{%
}{.}\hspace{.4pt}1145\discretionary{/}{%
}{/}1361192\hspace{.1pt}\discretionary{.}{%
}{.}\hspace{.4pt}1361195}}


\bibitem{lai1992fortran}
M.-J. Lai.
\newblock Fortran subroutines for b-nets of box splines on three-and
  four-directional meshes.
\newblock {\em Numerical Algorithms}, 2(1):33--38, 1992.

\bibitem{laine2020modular}
S.~Laine, J.~Hellsten, T.~Karras, Y.~Seol, J.~Lehtinen, and T.~Aila.
\newblock Modular primitives for high-performance differentiable rendering.
\newblock {\em ACM Transactions on Graphics (TOG)}, 39(6):1--14, 2020.

\bibitem{lavoue2005subdivision}
G.~Lavou{\'e}, F.~Dupont, and A.~Baskurt.
\newblock Subdivision surface fitting for efficient compression and coding of
  3d models.
\newblock In {\em Visual Communications and Image Processing 2005}, vol. 5960,
  pp. 1159--1170. SPIE, 2005.

\bibitem{li2009robust}
H.~Li, B.~Adams, L.~J. Guibas, and M.~Pauly.
\newblock Robust single-view geometry and motion reconstruction.
\newblock {\em ACM Transactions on Graphics (ToG)}, 28(5):1--10, 2009.

\bibitem{litke2001fitting}
N.~Litke, A.~Levin, and P.~Schroder.
\newblock Fitting subdivision surfaces.
\newblock In {\em Proceedings Visualization, 2001. VIS'01.}, pp. 319--568.
  IEEE, 2001.

\bibitem{Liu2021MLS}
S.-L. Liu, H.-X. Guo, H.~Pan, P.-S. Wang, X.~Tong, and Y.~Liu.
\newblock Deep implicit moving least-squares functions for 3d reconstruction.
\newblock In {\em Proceedings of the IEEE/CVF Conference on Computer Vision and
  Pattern Recognition}, pp. 1788--1797, 2021.

\bibitem{Loop:1987}
C.~Loop.
\newblock Smooth subdivision surfaces based on triangles.
\newblock Master's thesis, University of Utah, USA, 1987.

\bibitem{Ma}
W.~Ma, X.~Ma, S.-K. Tso, and Z.~Pan.
\newblock Subdivision surface fitting from a dense triangle mesh.
\newblock In {\em Geometric Modeling and Processing. Theory and Applications.
  GMP 2002. Proceedings}, pp. 94--103, 2002. doi: {{%
10\hspace{.1pt}\discretionary{.}{%
}{.}\hspace{.4pt}1109\discretionary{/}{%
}{/}GMAP\hspace{.1pt}\discretionary{.}{%
}{.}\hspace{.4pt}2002\hspace{.1pt}\discretionary{.}{%
}{.}\hspace{.4pt}1027500}}


\bibitem{marinov2005optimization}
M.~Marinov and L.~Kobbelt.
\newblock Optimization methods for scattered data approximation with
  subdivision surfaces.
\newblock {\em Graphical Models}, 67(5):452--473, 2005.

\bibitem{mendhurwar2020system}
K.~Mendhurwar, G.~Handa, L.~Zhu, S.~Mudur, E.~Beauchesne, M.~LeVangie,
  A.~Hallihan, A.~Javadtalab, and T.~Popa.
\newblock A system for acquisition and modelling of ice-hockey stick shape
  deformation from player shot videos.
\newblock In {\em Proceedings of the IEEE/CVF Conference on Computer Vision and
  Pattern Recognition Workshops}, pp. 890--891, 2020.

\bibitem{meyer2003discrete}
M.~Meyer, M.~Desbrun, P.~Schr{\"o}der, and A.~H. Barr.
\newblock Discrete differential-geometry operators for triangulated
  2-manifolds.
\newblock In {\em Visualization and mathematics III}, pp. 35--57. Springer,
  2003.

\bibitem{mfoggp_dyn_reg_07}
N.~J. Mitra, S.~Flory, M.~Ovsjanikov, N.~Gelfand, L.~Guibas, and H.~Pottmann.
\newblock Dynamic geometry registration.
\newblock In {\em Symposium on Geometry Processing}, pp. 173--182, 2007.

\bibitem{montes2020computational}
J.~Montes, B.~Thomaszewski, S.~Mudur, and T.~Popa.
\newblock Computational design of skintight clothing.
\newblock {\em ACM Transactions on Graphics (TOG)}, 39(4):105--1, 2020.

\bibitem{newcombe2015dynamicfusion}
R.~A. Newcombe, D.~Fox, and S.~M. Seitz.
\newblock Dynamicfusion: Reconstruction and tracking of non-rigid scenes in
  real-time.
\newblock In {\em Proceedings of the IEEE conference on computer vision and
  pattern recognition}, pp. 343--352, 2015.

\bibitem{oztireli2009feature}
A.~C. {\"O}ztireli, G.~Guennebaud, and M.~Gross.
\newblock Feature preserving point set surfaces based on non-linear kernel
  regression.
\newblock In {\em Computer graphics forum}, vol.~28, pp. 493--501. Wiley Online
  Library, 2009.

\bibitem{PiegTill96}
L.~Piegl and W.~Tiller.
\newblock {\em The NURBS Book}.
\newblock Springer-Verlag, New York, NY, USA, second ed., 1996.

\bibitem{popa2010globally}
T.~Popa, I.~South-Dickinson, D.~Bradley, A.~Sheffer, and W.~Heidrich.
\newblock Globally consistent space-time reconstruction.
\newblock In {\em Computer Graphics Forum}, vol.~29, pp. 1633--1642. Wiley
  Online Library, 2010.

\bibitem{ravi2020accelerating}
N.~Ravi, J.~Reizenstein, D.~Novotny, T.~Gordon, W.-Y. Lo, J.~Johnson, and
  G.~Gkioxari.
\newblock Accelerating 3d deep learning with pytorch3d.
\newblock {\em arXiv preprint arXiv:2007.08501}, 2020.

\bibitem{sharf2008space}
A.~Sharf, D.~A. Alcantara, T.~Lewiner, C.~Greif, A.~Sheffer, N.~Amenta, and
  D.~Cohen-Or.
\newblock Space-time surface reconstruction using incompressible flow.
\newblock {\em ACM Transactions on Graphics (TOG)}, 27(5):1--10, 2008.

\bibitem{sharma2020parsenet}
G.~Sharma, D.~Liu, S.~Maji, E.~Kalogerakis, S.~Chaudhuri, and R.~M{\v{e}}ch.
\newblock Parsenet: A parametric surface fitting network for 3d point clouds.
\newblock In {\em European Conference on Computer Vision}, pp. 261--276.
  Springer, 2020.

\bibitem{shinya2004unifying}
M.~Shinya.
\newblock Unifying measured point sequences of deforming objects.
\newblock In {\em Proceedings. 2nd International Symposium on 3D Data
  Processing, Visualization and Transmission, 2004. 3DPVT 2004.}, pp. 904--911.
  IEEE, 2004.

\bibitem{sorkine2007rigid}
O.~Sorkine and M.~Alexa.
\newblock As-rigid-as-possible surface modeling.
\newblock In {\em Symposium on Geometry processing}, vol.~4, pp. 109--116,
  2007.

\bibitem{stam1998evaluation}
J.~Stam.
\newblock Evaluation of loop subdivision surfaces.
\newblock In {\em SIGGRAPH’98 CDROM Proceedings}. Citeseer, 1998.

\bibitem{Stanford}
Stanford.
\newblock The stanford 3d scanning repository.
\newblock \url{http://graphics.stanford.edu/data/3Dscanrep/}.

\bibitem{stoll2006template}
C.~Stoll, Z.~Karni, C.~R{\"o}ssl, H.~Yamauchi, and H.-P. Seidel.
\newblock Template deformation for point cloud fitting.
\newblock In {\em PBG@ SIGGRAPH}, pp. 27--35, 2006.

\bibitem{sumner2004deformation}
R.~W. Sumner and J.~Popovi{\'c}.
\newblock Deformation transfer for triangle meshes.
\newblock {\em ACM Transactions on graphics (TOG)}, 23(3):399--405, 2004.

\bibitem{sussmuth2008reconstructing}
J.~S{\"u}{\ss}muth, M.~Winter, and G.~Greiner.
\newblock Reconstructing animated meshes from time-varying point clouds.
\newblock In {\em Proceedings of the Symposium on Geometry Processing}, pp.
  1469--1476, 2008.

\bibitem{tevs2012animation}
A.~Tevs, A.~Berner, M.~Wand, I.~Ihrke, M.~Bokeloh, J.~Kerber, and H.-P. Seidel.
\newblock Animation cartography—intrinsic reconstruction of shape and motion.
\newblock {\em ACM Transactions on Graphics (TOG)}, 31(2):1--15, 2012.

\bibitem{umeyama1991least}
S.~Umeyama.
\newblock Least-squares estimation of transformation parameters between two
  point patterns.
\newblock {\em IEEE Transactions on Pattern Analysis \& Machine Intelligence},
  13(04):376--380, 1991.

\bibitem{wand2009efficient}
M.~Wand, B.~Adams, M.~Ovsjanikov, A.~Berner, M.~Bokeloh, P.~Jenke, L.~Guibas,
  H.-P. Seidel, and A.~Schilling.
\newblock Efficient reconstruction of nonrigid shape and motion from real-time
  3d scanner data.
\newblock {\em ACM Transactions on Graphics (TOG)}, 28(2):1--15, 2009.

\bibitem{wang2006fitting}
W.~Wang, H.~Pottmann, and Y.~Liu.
\newblock Fitting b-spline curves to point clouds by curvature-based squared
  distance minimization.
\newblock {\em ACM Transactions on Graphics (ToG)}, 25(2):214--238, 2006.

\bibitem{yumer2012surface}
M.~E. Yumer and L.~B. Kara.
\newblock Surface creation on unstructured point sets using neural networks.
\newblock {\em Computer-Aided Design}, 44(7):644--656, 2012.

\bibitem{zheng2012fast}
W.~Zheng, P.~Bo, Y.~Liu, and W.~Wang.
\newblock Fast b-spline curve fitting by l-bfgs.
\newblock {\em Computer Aided Geometric Design}, 29(7):448--462, 2012.

\bibitem{zollhofer2014real}
M.~Zollh{\"o}fer, M.~Nie{\ss}ner, S.~Izadi, C.~Rehmann, C.~Zach, M.~Fisher,
  C.~Wu, A.~Fitzgibbon, C.~Loop, C.~Theobalt, et~al.
\newblock Real-time non-rigid reconstruction using an rgb-d camera.
\newblock {\em ACM Transactions on Graphics (ToG)}, 33(4):1--12, 2014.

\end{thebibliography}
\end{document}